\documentclass[
reprint,
 superscriptaddress,
amsmath, amssymb,
aps,
pra,
]{revtex4-2}
\usepackage{amsmath}
\usepackage{xcolor}
\usepackage{graphicx}
\usepackage{dcolumn}
\usepackage{bm}
\usepackage[colorlinks=true, allcolors=blue]{hyperref}
\graphicspath{{figures/}{figures_sm/}}
\usepackage[
separate-uncertainty=false,
multi-part-units=single,
per-mode=symbol,
]{siunitx}
\usepackage{braket}
\usepackage{xspace}

\DeclareMathOperator\erf{erf}
\usepackage{xstring}
\newcommand*{\aref}[1]{%
\IfBeginWith{#1}{eq:}{Eq.~\eqref{#1}}{}%
\IfBeginWith{#1}{fig:}{Fig.~\ref{#1}}{}%
\IfBeginWith{#1}{tab:}{Table~\ref{#1}}{}%
\IfBeginWith{#1}{appendix:}{Appendix~\ref{#1}}{}%
\IfBeginWith{#1}{sec:}{Section~\ref{#1}}{}%
}
\DeclareUnicodeCharacter{0080}{ }
\newcommand{\appropto}{\mathrel{\vcenter{
\offinterlineskip\halign{\hfil$##$\cr
\sim\cr\noalign{\kern2pt}\propto\cr\noalign{\kern-2pt}}}}}
    
\begin{document}
\title{Measurement of the order parameter and its spatial fluctuations \\ across Bose-Einstein condensation}

\date{\today}

\author{Louise Wolswijk}
\affiliation{INO-CNR BEC Center and Dipartimento di Fisica, Universit\`a di Trento, 38123 Povo, Italy}

\author{Carmelo Mordini}
\altaffiliation[Current address: Institute for Quantum Electronics, ETH Z\"urich, Otto-Stern-Weg 1, 8093 Z\"urich, Switzerland and Quantum Center, ETH Z\"urich, 8093 Z\"urich, Switzerland]{}
\affiliation{INO-CNR BEC Center and Dipartimento di Fisica, Universit\`a di Trento, 38123 Povo, Italy}

\author{Arturo Farolfi}
\affiliation{INO-CNR BEC Center and Dipartimento di Fisica, Universit\`a di Trento, 38123 Povo, Italy}

\author{Dimitris Trypogeorgos}
\altaffiliation[Current address: CNR Nanotec, Institute of Nanotechnology, via Monteroni, 73100, Lecce, Italy]{}
\affiliation{INO-CNR BEC Center and Dipartimento di Fisica, Universit\`a di Trento, 38123 Povo, Italy}

\author{Franco Dalfovo}
\affiliation{INO-CNR BEC Center and Dipartimento di Fisica, Universit\`a di Trento, 38123 Povo, Italy}

\author{Alessandro Zenesini}
\email[]{alessandro.zenesini@ino.cnr.it}
\author{Gabriele Ferrari}
\author{Giacomo Lamporesi}
\affiliation{INO-CNR BEC Center and Dipartimento di Fisica, Universit\`a di Trento, 38123 Povo, Italy}

\begin{abstract}
We investigate the strong out-of-equilibrium dynamics occurring when a harmonically trapped ultracold bosonic gas is evaporatively cooled across the Bose--Einstein condensation transition. 
By imaging the cloud after free expansion, we study how the cooling rate affects the timescales for the growth of the condensate order parameter and the relaxation dynamics of its spatial fluctuations. 
We find evidence of a delay on the condensate formation related to the collisional properties and a universal condensate growth following the cooling rate. Finally, we measure an exponential relaxation of the spatial fluctuations of the order parameter that also shows a universal scaling. Notably, the scaling for the condensate growth and for the relaxation of its fluctuations follow different power laws.
\end{abstract}

\maketitle

\section{Introduction}

The increasing relevance of quantum technologies demands an in-depth understanding and a higher degree of control of the constituents of a quantum system. In particular, quantum many-body systems are characterized by a rich variety of phases that can be achieved starting from a given initial condition \cite{Baumann2010,gutierrez2017,zhang2017,GrossBloch2017,kennes2018,kai2020,fontaine2021}. The capability to reach a well-known target state is directly connected to the ability of controlling the system throughout the transfer, not necessarily through equilibrium configurations, but also crossing strongly out-of-equilibrium ones. Whenever a phase transition is present along the path, the control becomes even more challenging  \cite{Bason2012,Richerme2013,choi2020,guo2021}. 
In particular, the crossing of a phase transition attracts a lot of interest for what concerns the search for universality classes defined by common critical exponents that describe the scaling laws of the observables.
The Kibble--Zurek mechanism represents a milestone result \cite{Kibble76,Zurek85}
in the general description of systems quenched across a second-order phase transition accompanied by a symmetry breaking.
It describes how an extended system can only locally break the symmetry and form causally disconnected domains of the order parameter, whose number and size 
scale with the cooling rate following a universal power law.
First experiments testing the Kibble--Zurek mechanism were performed in superfluid Helium \cite{Bauerle96}, and later on also in superconductors \cite{Carmi99, Monaco06} and trapped ions \cite{Pyka13,Ulm13}.

In the last years, thanks to the exquisite control experimentally available on ultracold atoms, one of the most studied second-order phase transitions is Bose-Einstein condensation (BEC) \cite{Anderson95, Davis95, Dalfovo90}.
The formation of an order parameter within the initial thermal atomic cloud represents a paradigmatic test-bench where to study the effects of cooling ramps crossing the critical point at variable rate, by means of a large control on experimental parameters \cite{Ensher96,Miesner98,Kohl02,Eigen2018,Glidden21}.
Further experiments in ultracold atomic gases focused on the emergence of coherence and the statistics of the formation of topological defects \cite{Lamporesi13, Corman14, Braun15, Chomaz15, Navon15}.

The Kibble-Zurek mechanism, however, does not fully describe the out-of-equilibrium dynamics of the system at the transition, nor the post-quench interaction mechanism between different domains leading to coarse graining. Most of the theoretical models consider a linear variation in time of a single parameter across the critical point and approximate the quench with adiabatic (far from the transition) or frozen (across the transition) regimes \cite{delCampo14}.
In real experiments, the variation of the control parameter --- usually the temperature of the ultracold atomic sample --- is accompanied by the variation of other quantities such as the atom number or the collisional rate, making it more difficult to accurately describe the system across the transition and to predict its properties after the quench.
Recent works studied the condensate formation mechanism including effects such as inhomogeneity, atom number decay and defect number saturation \cite{delCampo13, Lamporesi13,Liu18,Liu20,Donadello16,Goo21} going beyond the Kibble-Zurek mechanism.

In this work, we present an experimental investigation of the dynamics of the order parameter and its spatial fluctuations in an ultracold bosonic gas as a function of the cooling rate across the BEC transition point.
We study the relevant timescales associated to such dynamics
and identify universal power-law scaling. 

In \aref{sec:ExpProc}, we describe the experimental sequence and the procedure to extract the relevant observables. \aref{sec:Growth} reports on the analysis of the condensate growth and its dependency on the cooling rate and the initial atom number.
In \aref{sec:Relaxation}, we discuss the dynamics of the phase fluctuations and their relaxation toward the equilibrium state. Concluding remarks are reported in \aref{sec:Conclusion}.

\section{The experimental procedure}
\label{sec:ExpProc}

\begin{figure*}[t]
 \centering
 \includegraphics[width=1.55\columnwidth]{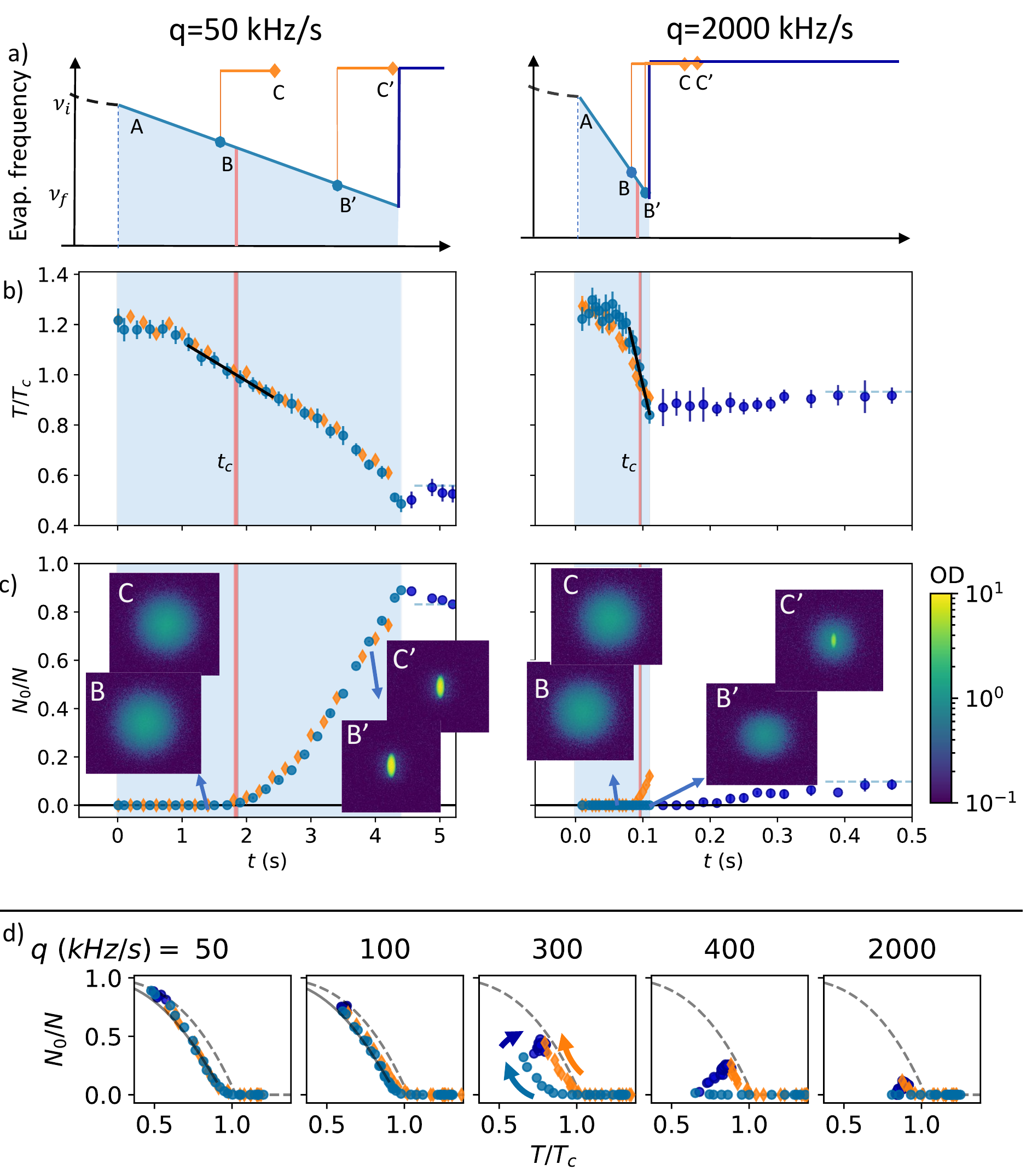}
 \caption{Exploring the gas parameters during the evaporation ramp for a low (left column) or a high (right) cooling rate. (a) Variation of the evaporation frequency $\nu$ in time; the evaporation rate is $q=d\nu/dt$. Temperature (b) and condensed fraction (c) measured by interrupting the evaporation ramp at different times $t$ and imaging the cloud either immediately (light and dark blue circles) or keeping the gas in the trap for equilibration (orange diamonds). Temperature is extracted from a Bose distribution fit to the tails of the atomic distribution measured after time of flight and is normalized to the transition temperature $T_c$. The vertical red line represents the time $t_c$ at which $T_c$ is experimentally crossed; the horizontal blue dashed line is the final temperature after equilibration. The horizontal dashed line in (c) marks the final condensate fraction. Insets show the absorption images taken in points B, C, B', C'. The shaded area marks the duration of the ramp. (d) Evolution of $N_0/N$ versus $T/T_c$, showing that higher evaporation rates induce strong out-of-equilibrium dynamics, as can be seen comparing the instantaneous measurements (light and dark blue circles) to the corresponding equilibrium points (orange diamonds). The arrows indicate the direction of the time-evolution. The dashed and the solid lines represent the theoretical predictions for the condensate fraction for an ideal and for an interacting Bose gas, respectively. 
 }
 \label{fig:fig1}
\end{figure*}

A gas of sodium atoms is initially laser cooled and then transferred to an elongated harmonic magnetic trap that has a cylindrical symmetry ($\omega_r=\omega_y=\omega_z$) around the horizontal $x$ axis. The trapping frequencies are $\omega_x / 2\pi = \SI{12.30 \pm 0.05}{\hertz}$ and $\omega_r / 2\pi = \SI{138 \pm 1 }{\hertz}$. During the evaporation, the bias  magnetic field is 1.53 G.
A first evaporative cooling stage is performed by exponentially sweeping the frequency of a radio-frequency field that transfers $|F,m_F\rangle=|1,-1\rangle$ trapped atoms to the non trapped $|1,0\rangle$ state. At the end of this stage [point A in \aref{fig:fig1}(a)], the trap has a depth of about $k_B \times \SI{12}{\micro\kelvin}$.

At this point, a second evaporative stage, relevant for the analysis reported in this paper, begins.
During this second cooling stage (shaded region in \aref{fig:fig1}), we apply a linear ramp of the evaporation frequency with rate $q=d\nu/dt$, down to a final frequency $\nu_f$, which is reached at time $t_{f}$. This final frequency lies enough above the trap bottom value to avoid the removal of atoms from the condensate and the creation of collective excitations. 

We image the atomic cloud with calibrated resonant absorption imaging \cite{Mordini20} along the horizontal $y$ direction, after 
a time of flight (TOF) from the trap switch--off of $t_\text{TOF}=\SI{50}{\milli\second}$. We interleave two different measurement schemes: i) \textit{instantaneous} images, where we probe the status of the sample during the linear cooling ramp, at different times $t$ from the ramp beginning, by interrupting the evaporation and immediately releasing the atoms (B, B'); ii) \textit{equilibrium} images, where we interrupt the evaporation at the same time $t$, but leave the condensate in the trap for an extra hold time of 1 s before releasing (C, C').
During the hold time, we set the evaporation frequency to $\nu_\text{hold} = \nu_i+\SI{400}{\kilo\hertz}$, corresponding to about \SI{31}{\micro\kelvin}, to ensure thermalization without further atom loss.
In both cases, we fit the outer part of the atomic cloud with a Bose function to measure the width and amplitude of the distribution, from which we extract the atom number in the thermal component. Subtracting the fitted Bose function from the data, we obtain an image of the condensate, that we integrate and obtain the number of atoms in the condensate, $N_0$.
We do not fit the condensate with a Thomas--Fermi profile since it properly describes the condensate profile only when the system is in its ground state and not in the presence of fluctuations.
The total atom number $N$ is then given by the sum of $N_0$ and the number of atoms in the thermal part.

The evaporation process removes the most energetic atoms from the distribution and leads to a decrease in the total atom number. Therefore, the gas is no longer in thermal equilibrium, especially for the fastest ramps.
For this reason, we introduce an effective temperature $T$  according to $T = m \omega_z^2  \sigma_z^2 / [(1 + \omega_z^2 t_\text{TOF}^2)k_B]$, where $m$ is the atomic mass and $\sigma_z$ the width of the Bose distribution 
measured along the radial direction. Given that $\omega_z\gg\omega_x$ and that the condition $\omega_z t_\text{TOF}\gg1$ is satisfied, the distribution along the $z$ direction provides indeed a more reliable estimate for the temperature than the one along the axial direction $x$.

Figure~\ref{fig:fig1}(b) shows the temperature during the evaporation in units of the critical temperature for condensation,  $T_c$. At each time $t$, we estimate the corresponding $T_c$ considering the measured instantaneous atom number $N$ at such time. 
We define $T_c=T_c^0 + \delta T_c$, where
\begin{equation}
T_c^0 = 0.94\,\frac{\hbar \omega_{\rm ho}}{k_B} N^{1/3}
\end{equation}
 is the critical temperature for an ideal gas in the same trap, with $k_B$ the Boltzmann constant and $\omega_{\rm ho}= (\omega_x \omega_r^2)^{1/3}$, while $\delta T_c$ is a correction due to interaction effects. The latter is estimated to be
 \begin{equation}
\delta T_c=-1.3 \frac{a}{a_{\rm ho}} N^{1/6}
\end{equation}
 as in Eq.~(122) of Ref.\,\cite{Dalfovo99}, where $a$ is the $s$-wave scattering length and $a_{\rm ho}= \sqrt{\hbar/m\omega_{\rm ho}}$. In our case, such a correction is $|\delta T_c|/T_c^0\sim 3\%$. We instead neglect finite size corrections [see Eq.~(20) in \cite{Dalfovo99}] that are about one order of magnitude smaller.

Light blue circles correspond to instantaneous images taken before the end of the linear evaporation ramp, while orange diamonds are the corresponding equilibrium images.
Dark blue circles are instantaneous pictures taken after the end of the ramp, where we raise the evaporation frequency to $\nu_\text{hold}$ and keep the condensate in trap to probe the post-quench relaxation dynamics.
Each point is the average of up to 10 experimental shots, and error bars represent one standard deviation.
The red vertical line marks the critical time $t_c$ at which $T/T_c = 1$, obtained by means of a linear fit of the data in the region around the critical point (solid black line). From the fit, we also extract the characteristic quench time
\begin{equation}
\tau_Q=\frac{T_c}{|(dT/dt)_{t_c}|},  
\end{equation}
which allows for a direct comparison to the theoretical models assuming a linear variation of the temperature across the transition.

The condensate fraction $N_0 / N$ is shown in \aref{fig:fig1}(c), again comparing the instantaneous and the equilibrium configurations. The insets show examples of TOF images before and after the transition point.
For a sufficiently slow ramp [left panels in \aref{fig:fig1}(a-c)], the condensate forms during the evaporation ramp, whereas for a fast ramp, no condensate fraction is present at the end of the ramp, but it emerges afterwards, on a time scale independent of the evaporation rate $q$, but rather intrinsic of the system.

In \aref{fig:fig1}(d), we show the dynamical evolution of the gas across the transition, by plotting $N_0 / N$ as a function of $T / T_c$ for different $q$. 
This highlights the out-of-equilibrium evolution related to the cooling process. While for the smallest evaporation rates the gas cools down almost adiabatically, as pointed out by the overlapping instantaneous and equilibrium curves for $q=\SI{50}{\kilo\hertz/\s}$, for faster evaporation rates the gas is driven in an increasingly out-of-equilibrium state, as shown for instance in the $q=\SI{300}{\kilo\hertz/\s}$ panel, where the arrows indicate the direction of the time evolution of the instantaneous images (blue and dark blue) and the equilibrium ones (orange). 
We observe that the instantaneous state of the gas strongly diverges from the equilibrium one while driven by evaporation, and it relaxes towards the final equilibrium state after the end of the ramp.
We checked that, even for the highest evaporation rates ($q$ up to $\SI{4000}{\kilo\hertz/\s}$), at any time $t \gtrsim \SI{500}{ms}$ after the end of the ramp, most of the excitations have relaxed and the system reaches a stationary state. For this reason, we can consider the images taken after $\SI{1}{s}$ waiting time as reliable equilibrium points. For comparison, in \aref{fig:fig1}(d) we also show the theoretical prediction for the ideal Bose gas (dashed line)
\begin{equation}
 \frac{N_0}{N}=1-\left( \frac{T}{T_c^0}\right)^3,   
\end{equation}
and for the interacting Bose gas (solid line) \cite{Giorgini97,Dalfovo99}
\begin{equation}
 \frac{N_0}{N}=1-\left( \frac{T}{T_c^0}\right)^3 - \frac{\zeta(2)}{\zeta(3)}\eta  \left(\frac{T}{T_c^0}\right)^2 \left[1-\left(\frac{T}{T_c^0}\right)^3\right]^{\frac{2}{5}}
 \label{eq:non}
\end{equation}
where $\zeta(n)$ is the Riemann function and $\eta=\mu/kT_c^0$, $\mu$ being the gas chemical potential. Expression (\ref{eq:non}) is predicted to fail close to $T_c$ and, for this reason, we plot the corresponding curve only in the first two panels of \aref{fig:fig1}(d), where experimental data are available at low temperature.  

\begin{figure*}
 \centering
\includegraphics[width=2\columnwidth]{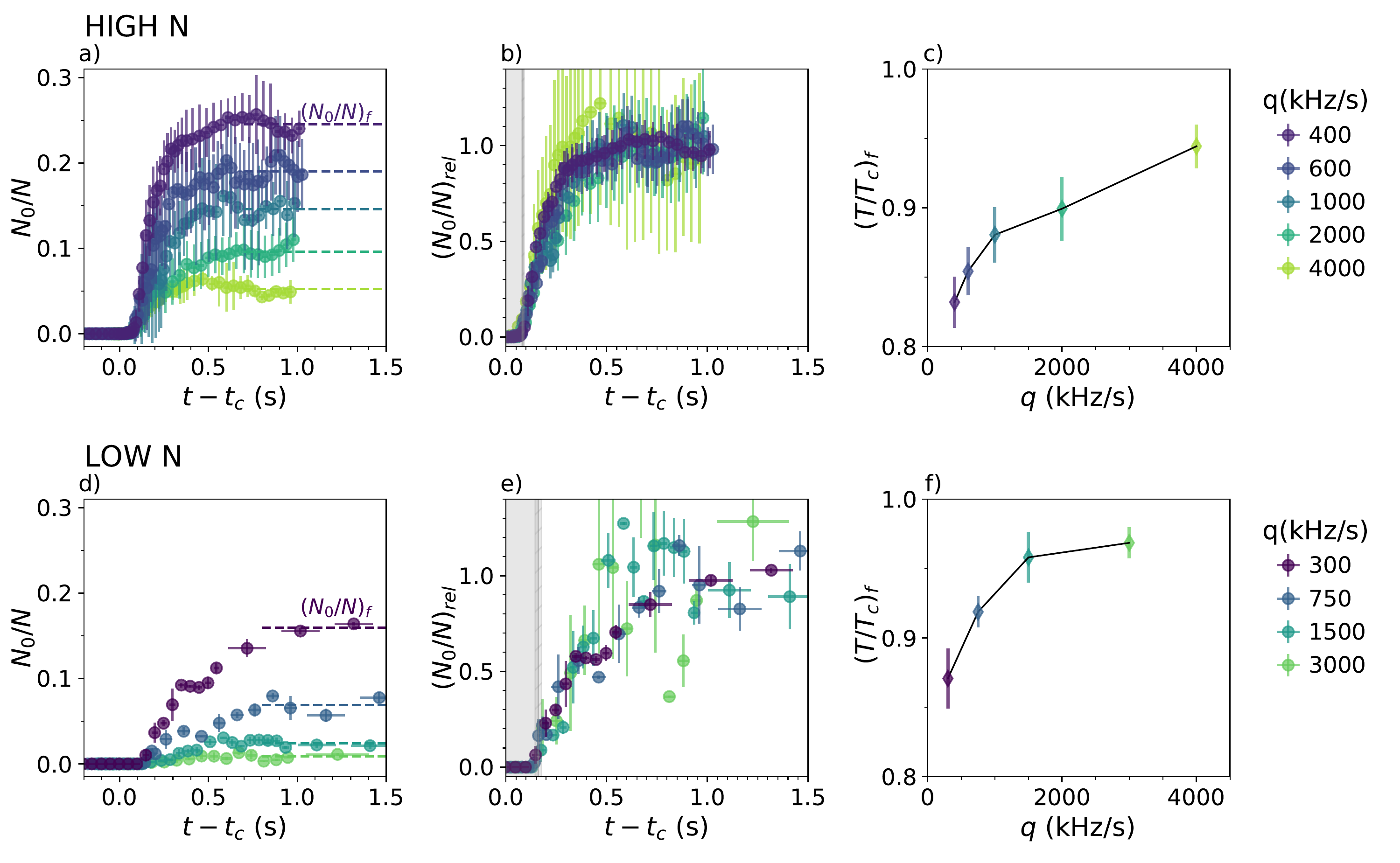}
 \caption{Post-quench growth dynamics of the BEC order parameter for high evaporation rates. (a) Condensate fraction as a function of time for atom number $N_\text{high}$, for different values of $q$. The time axes are shifted according to $t_c$, see text. The dashed lines mark the equilibrium condensed fraction for each evaporation rate. (b) Same as (a), but with the vertical axis normalized to the final condensed fraction for each evaporation rate.  The gray area highlights the latency time $\Delta t$. (c) Final relative temperature for the various rates $q$.
 (d-f) Equivalent results for a smaller atom number, $N_\text{low}$.  }
 \label{fig:fig2}
\end{figure*}

\section{Growth of the order parameter}
\label{sec:Growth}
In this Section we look closely at the dynamical formation of the condensate order parameter, following the evaporative cooling. 

In all the performed experiments, while the critical point $t_c$ always lies within the linear evaporation ramp, the time evolution of the condensate depends on the evaporation rate and shows two distinct regimes, that we identify as high and low evaporation rates.

For the high evaporation rates, condensation occurs after the end of the linear ramp, driven by thermalization at constant atom number. 
For both regimes, we explore the dependence on the initial atom number by preparing ensembles in two different initial conditions (measured at point A of the experimental sequence), one with high atom number $N_\text{high} = \num{3.6(3)e7}$ and initial temperature $T_\text{high}= \SI{1.14(3)}{\micro\kelvin} = 1.28(7) T_c$, and one with lower atom number $N_\text{low} = \num{1.70(9)e7}$ and temperature $T_\text{low} = \SI{0.92(4)}{\micro\kelvin} = 1.32(5) T_c$.

\subsection{High evaporation rate}
\label{sec:fast_ramps}

In \aref{fig:fig2}(a), we show the growth of the condensate fraction in samples with $N_\text{high}$, for high evaporation rates.
Although the critical temperature is crossed at time $t_c$ during the ramp, the order parameter forms later, after an average \textit{latency time} (in literature also referred to as  \textit{initiation time} \cite{Kohl02})  $\Delta t = 85(7)$ ms past $t_c$. 
The final BEC fraction $(N_0/N)_f$ [obtained by averaging the data for $(t - t_c) > \SI{500}{ms}$ and marked by the dashed lines in \aref{fig:fig2}(a)] is then slowly reached on a time scale that appears to be independent of $q$. This proofs that these ramps are effectively a quasi-instantaneous quench. 
The final temperature $(T/T_c)_f$ increases with $q$, direct consequence of a less and less efficient evaporation, as shown in \aref{fig:fig2}(c). In \aref{fig:fig2}(b), the points are vertically rescaled by $(N_0/N)_f$. The rescaling operation highlights that, for all the evaporation rates, the onset of the order parameter occurs at the same time and the rescaled condensate fraction evolution follows the same dynamics.

\begin{figure*}
 \centering
\includegraphics[width=2.0\columnwidth]{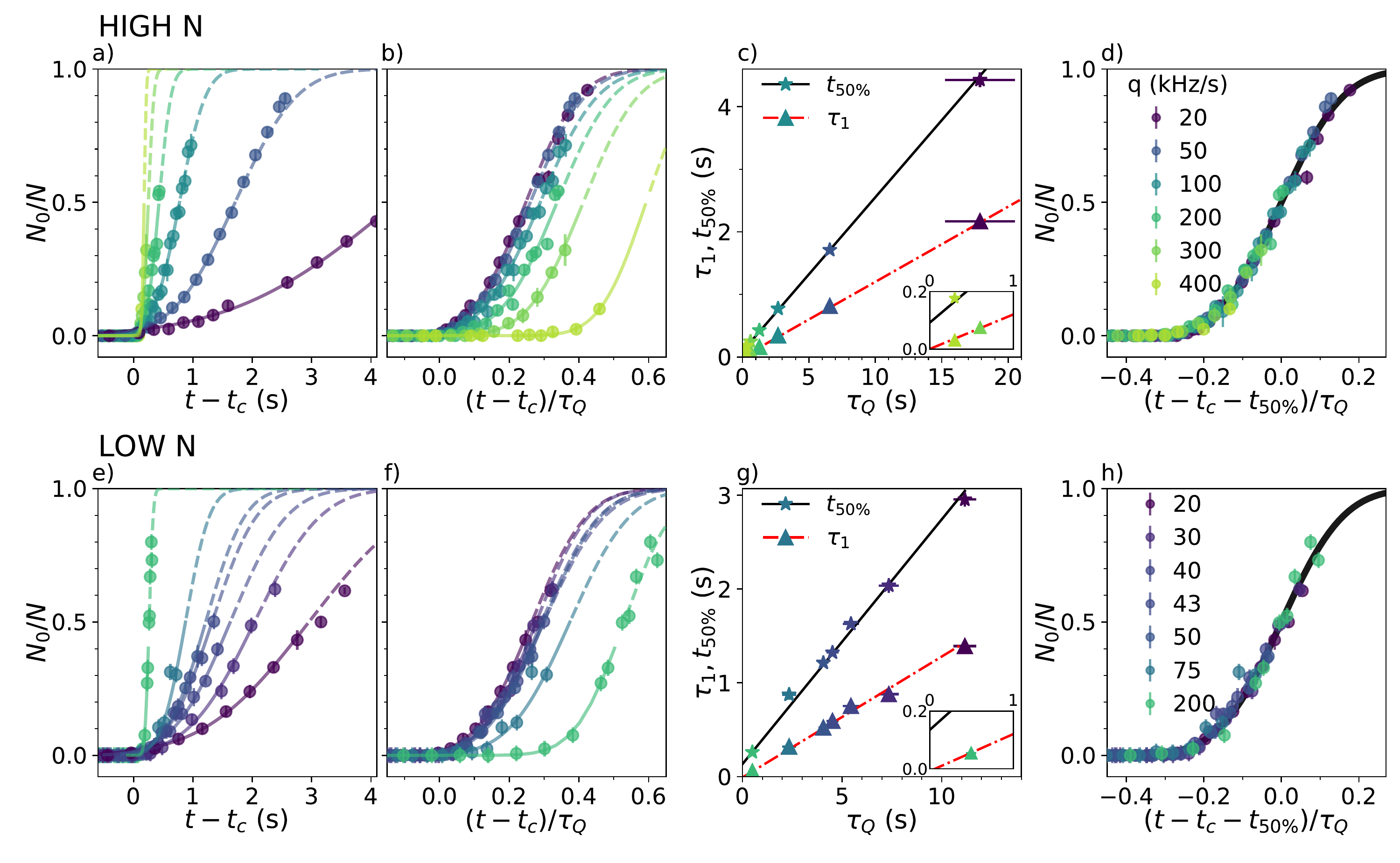}
 \caption{
 Growth of the BEC order parameter during cooling ramps with low evaporation rate. The condensate fraction is shown for different evaporation rates on gases with $N_\text{high}$ atoms, as a function of $(t-t_c)$ (a) or of $(t-t_c)/\tau_Q$ (b). Panel (c) shows how $\tau_1$ (triangles) and $t_{50\%}$ (stars) depend on $\tau_Q$. The lines are linear fits of $\tau_1(\tau_Q)$ (solid) and $t_{50\%}(\tau_Q)$ (dot-dashed). The inset shows a zoom of the region close to the origin, highlighting the presence of a finite offset for $t_{50\%}$, while $\tau_1$ tends to zero when $\tau_Q\rightarrow0$.  In Panel (d) all growth curves overlap onto a universal growth curve (in black) when shifted by the $t_{50\%}$ parameter obtained from the fit and rescaled by the time $\tau_{Q}$ of each ramp. (e-h) Same as (a-d), but for a smaller initial atom number, $N_\text{low}.$ The black curve in panels (h) is the same as in panel (d), showing that the universality holds also for different atom numbers. 
}
 \label{fig:fig3}
\end{figure*}

Figures~\ref{fig:fig2}(d-f) show analogous results for the dataset with $N_\text{low}$.
Qualitatively, the behavior is the same as in the high atom number case, but the delay in the onset of the order parameter becomes larger, $\Delta t = 161(16)$ ms. 
Similarly, the growth of the condensate fraction is slower for less dense clouds,  with the final equilibrium condensed fraction being reached after $(t-t_c)>\SI{750}{\milli\second}$.
Both behaviors well agree with a larger collisional time in the case of lower atomic density. Estimating the classical collisional time as  $\tau_\text{coll}=(\bar{n}\sigma v)^{-1}$ \cite{Dalibard99}, where $\bar{n}=\int{n^2(r) dr}/\int{n(r) dr}$ is the average density of the thermal cloud at the transition, $\sigma=8\pi a^2$ the scattering cross-section and $v=4\sqrt{k_B T/(\pi m)}$ the average velocity, we have $\tau_\text{coll}\sim\SI{18}{ms}$ for the dataset with $N_\text{high}$, and $\tau_\text{coll}\sim\SI{32}{ms}$ for $N_\text{low}$.
Therefore, we find that the time delay for the onset of condensation also obeys a universal behavior, resulting in $\Delta t \sim 5\,\tau_\text{coll}$.
These observations are in good agreement with early works on the condensation process after a sudden quench \cite{Miesner98,Kohl02} which clarified the role of interactions throughout a complete quantum kinetic theory \cite{Gardiner1997} and by pointing up the existence of a \textit{latency time} in the condensation onset. 

\subsection{Low evaporation rate}
\label{sec:slow_ramps}

The mechanism discussed in \aref{sec:fast_ramps} deals with almost instantaneous quenches and investigates the post-quench growth of the BEC order parameter.

In this Section, instead, we study the case of lower evaporation rates, with $q\leq\SI{400}{\kilo\hertz\per\second}$ in the case of $N_\text{high}$ and up to \SI{200}{\kilo\hertz\per\second} in the case of $N_\text{low}$. 
The dynamics is now conceptually different since the condensate formation occurs while the evaporation ramp is on, implying that the atom number and the temperature are both decreasing.

In \aref{fig:fig3}, we show the evolution of the condensate fraction as a function of $t-t_c$. The data for different $q$ [\aref{fig:fig3}(a)] fit well to a smooth step function. We choose the error function
\begin{equation}
\frac{N_0}{N}(t)= \frac{1}{2} \left[ 1 + \erf \left( \frac{t-t_c-t_{50\%}}{\sqrt{2}\ \tau_1} \right) \right],
\end{equation} 
centered at $t_{50\%}$ and with a growth time constant $\tau_1$. The fit is performed only on the data points corresponding to $N_0 / N < 0.5$, since we focus on the initial part of the condensate growth, and not on the saturation of the condensate fraction.

As expected, the growth is slower for smaller values of $q$ (higher $\tau_Q$). Figure~\ref{fig:fig3}(b) shows the same data as a function of the normalized time $(t - t_c) / \tau_Q$. In these units, the curves have the same growth rate and are relatively shifted in time, highlighting the fact that $\tau_Q$ is the timescale that dominates the process of condensate formation.

In Fig.~\ref{fig:fig3}(c), we report the fit results for $\tau_1$ and $t_{50\%}$. One can clearly see that both parameters scale linearly with $\tau_Q$. However, $t_{50\%}$ does not go to zero for $\tau_Q\rightarrow 0$, but instead to a finite value of the order of $\Delta t$.
In fact, this is consistent with the presence of a minimum latency time that the system requires before exhibiting the onset of condensation after the temperature has crossed the critical value, as already observed in the case of high evaporation rates. 
The same analysis is done for $N_\text{low}$ and reported in Fig.~\ref{fig:fig3}(e-g). In particular, we notice again that the delay decreases when the atom number increases. The characteristic time $\tau_1$ appears instead to be proportional to $\tau_Q$, with a similar proportionality factor for both regimes of atom numbers. The results of our fits are reported in Table~\ref{table:tab1}.

\begin{table}[b]
\begin{center}
\begin{tabular}{ c  c }
\hline
\hline
 HIGH N &  \\ 
 \hline
 $\tau_1$ & $0.120(5)\tau_Q-1(4)$ ms  \\ 
 $t_{50\%}$ & $0.246(4)\tau_Q+92(3)$ ms \\
 \hline
 \hline
 LOW N & \\
 \hline
 $\tau_1$ & $0.128(4)\tau_Q-8(6)$ ms \\
 $t_{50\%}$ & $0.259(9)\tau_Q+135(9)$ ms \\
 \hline
 \hline
\end{tabular}
\end{center}
\caption{Results for $\tau_1$ and $t_{50\%}$, for high and low $N$, corresponding to the parameters of the linear fits reported in panels (c) and (g) of Fig.\,\ref{fig:fig3},  respectively.}
\label{table:tab1}
\end{table}

Finally, in Panels (d) and (h) of \aref{fig:fig3}, we plot the data as a function of the dimensionless \textit{universal time} $(t-t_c-t_{50\%})/\tau_Q$.
Thanks to the observed linear dependence of both $\tau_1$ and $t_{50\%}$ on $\tau_Q$, all curves collapse on each other in any other point indicating the presence of universality. Note that the growth curve is the same for both atom number ranges explored. In fact, the black line in \aref{fig:fig3}h), that corresponds to the best fit for the $N_\text{high}$ data in \aref{fig:fig3}d), perfectly overlaps also on the data with $N_\text{low}$.  This result cannot be simply interpreted in terms of instantaneous thermal equilibrium, because the overlap occurs for ramp rates in a broad range, from the quasi-adiabatic evaporation rates (e.g. $q\sim\SI{20}{\kilo\hertz/s}$) to the higher ones ($q\sim\SI{300}{\kilo\hertz/s}$), where the condensate formation clearly occurs in an out-of-equilibrium condition, as observed in \aref{fig:fig1}(d).

As an intermediate conclusion, for low evaporation rates, we notice that in our system the condensate formation is characterized, during the evaporation, by a universal exponential growth occurring on a timescale  $\tau_1 \simeq 0.125\,\tau_Q$. It is also worth mentioning that the theoretical analysis of Ref.~\cite{Liu20} pointed out the role of another timescale, i.e., the freeze-out time $\hat{t}$, which is predicted to affect the growth of the condensate at the early stage of formation in the Kibble-Zurek mechanism. Our experimental observations, however, seem not to provide evidences of $\hat{t}$, likely due to the effects of the latency time associated to the finite collisional time in our samples, which is not included in \cite{Liu20}.

\section{Relaxation dynamics of the fluctuations}
\label{sec:Relaxation}

\begin{figure}[t!]
 \centering
\includegraphics[width=1.0\columnwidth]{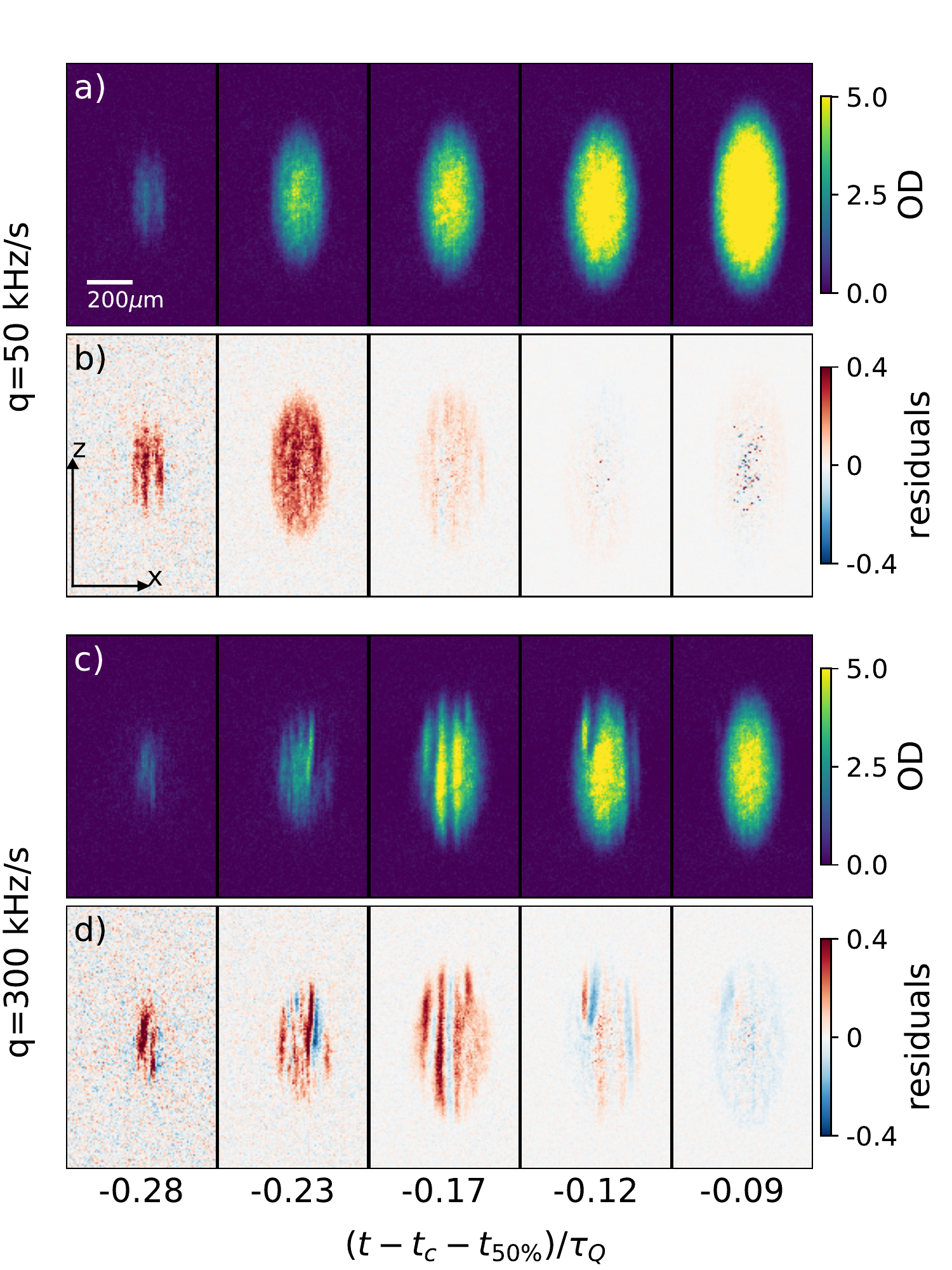}
 \caption{Relaxation dynamics of the order parameter fluctuations. (a) Pictures of the  BEC component as a function of the universal timescale of the condensate growth. (b) The residuals $h(x,z)$ between the OD shown in (a) and the average $\overline{OD}$ at the same universal time, normalized by  the peak $\overline{OD}$ at each time, are larger shortly after the BEC formation and then decay when the turbulence relaxes to the smooth equilibrium state. (a-b) and (c-d) correspond to evaporation rates of $q=\SI{50}{\kilo\hertz\per\second}$ and $q=\SI{300}{\kilo\hertz\per\second}$, respectively. 
 All images share the same spatial scale and coordinate axes indicated in a) and b).}
 \label{fig:fig4}
\end{figure}

In the previous Section, we limited our investigation to the study of the typical timescales characterizing the growth of the BEC order parameter in connection to the cooling rate. 
Now we take a closer look at the spatial fluctuations of the order parameter itself. In fact, when the condensate forms, an initially turbulent regime characterizes the whole system with strong spatial phase fluctuations \cite{Liu18}. Then a coarse-graining dynamics allows for the formation of domains that eventually turn into isolated quantized vortices. 

Phase fluctuations can be strongly enhanced in quantum gases for different reasons. Among them, reduced dimensionality and fast mechanical perturbations of the trapping potential have been widely studied both theoretically and experimentally.  

When the effective dimensionality of the gas is reduced, the long-range coherence is limited, enhancing phase fluctuations.
Bose gases in 1D have been studied by taking absorption images after a ballistic expansion of the cloud, since phase fluctuations present in situ evolve into density ripples after a long enough TOF \cite{Dettmer01,Imambekov09,Fabbri11}.
Bragg spectroscopy techniques were also used, connecting the width of the momentum distribution to the average length of the original phase domains \cite{Richard03,Fabbri11}.  
Quantum turbulence in ultracold gases \cite{Parker17} was recently studied by mechanically introducing it with rapid perturbations of the trapping optical or magnetic potentials \cite{Yukalov02,Parker05,Henn09,Neely13,Kwon14,Navon16,Tsatsos16,Groszek2016} and observing its relaxation phenomena. In these cases, the turbulent regime was explored using interferometric techniques or through the direct detection and counting of the generated quantized vortices.

Temperature quenches across the BEC critical point can also introduce strong phase fluctuations. 
The generated turbulent regime typically relaxes toward the BEC ground state, sometimes leaving isolated topological excitations, such as vortices in flat systems \cite{Weiler08,Goo21} or transverse solitonic vortices in elongated ones \cite{Ku14, Donadello14, Tylutki15}.
These defects, can be clearly observed on timescales of at least a few hundred of ms and were used in previous experiments to investigate the power-law scaling associated with the Kibble-Zurek mechanism \cite{Lamporesi13,Donadello16,Goo21}.  Here, we focus on the observation of the early-time turbulent phase generated by rapid cooling, and identify its typical relaxation timescale. 

In order to study the fluctuations of the order parameter, we image the gas after TOF and subtract the thermal component  resulting from the Bose function fit on the outer region. 
Figure\,\ref{fig:fig4} shows the optical density (OD) of the sole condensate component of gases released at different times shortly after the onset of condensation and during the BEC growth. The time frame is the universal one for the condensate growth, found in \aref{sec:slow_ramps}.
The upper [\aref{fig:fig4}(a)] and lower [\aref{fig:fig4}(c)] datasets
refer to a low and high evaporation rate, respectively.
Density ripples, similar to the ones observed in expanded 1D systems in Ref. \cite{Dettmer01}, are clearly visible. The difference, here, is that they are caused by the in-situ turbulent phase originating from finite-rate cooling.

Each BEC profile is compared to the average over several repetitions in the same experimental conditions. 
Fig.\ref{fig:fig4}(b) and \aref{fig:fig4}(d) show the residuals $h(x,z;t)$, evaluated as the difference between the OD of each image in panels (a) and (c) and the average $\overline{OD}$ at the corresponding time, normalized by the peak $\overline{OD}$: $h(x,z)=(OD-\overline{OD})/\overline{OD}_{\text{max}}$ .
In the early stages after the appearance of the BEC, a large amount of fluctuations is present, both at large and small scale.  We observe that for the higher evaporation rates there is a larger amount of fluctuations, which gradually decay on a longer universal time scale compared to the lower evaporation rates.

\begin{figure}[t!]
 \centering
\includegraphics[width=1.0\columnwidth]{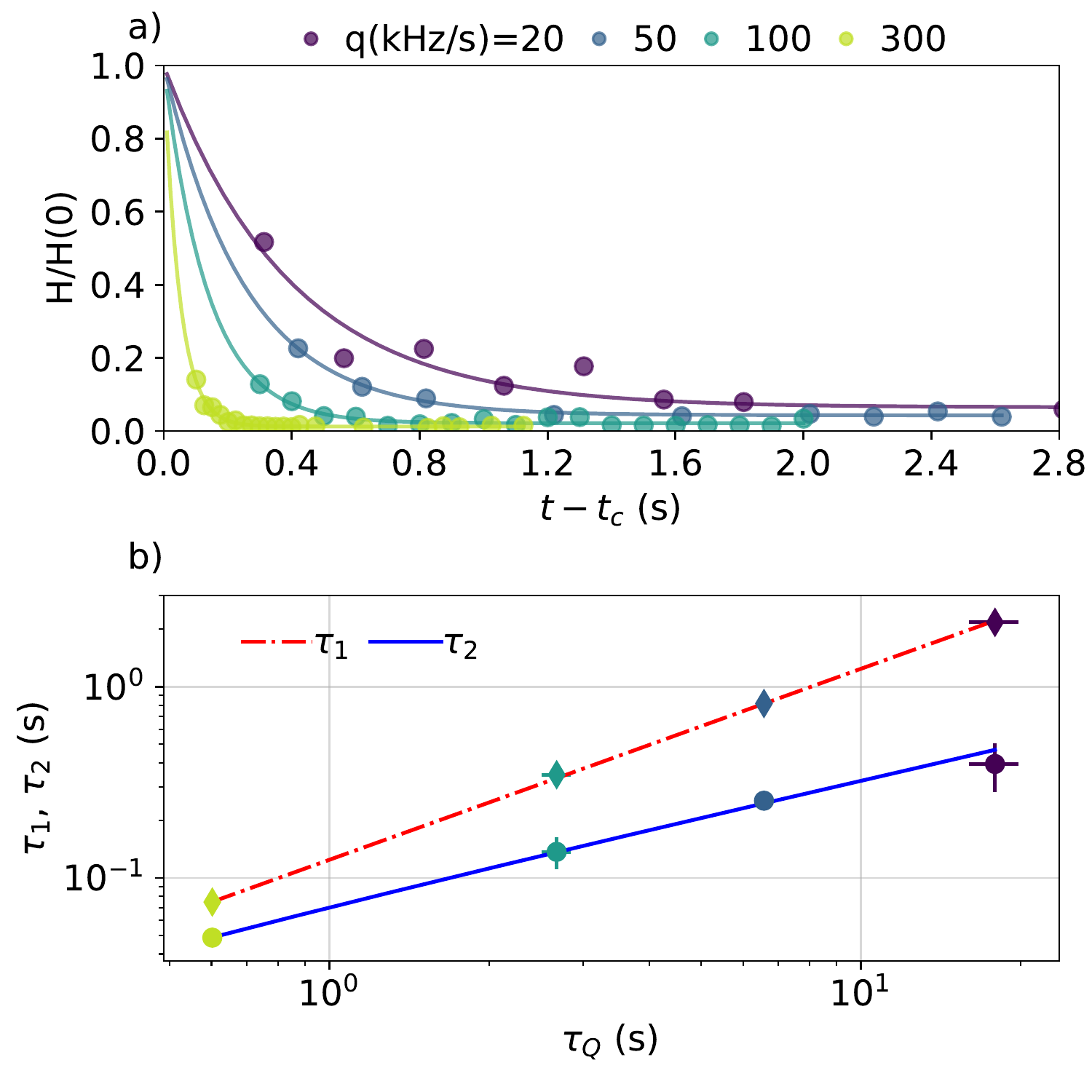}
 \caption{a) Decay of the order parameter fluctuations ${H(t-t_c)}$ in the case of  low evaporation rates, normalized by the initial value $H(0)$. b) Time constant $\tau_2$ of the exponential fit, compared with that of the condensate growth rate $\tau_1$. Linear fits in log-log scale highlight the power-law scaling behavior of the two quantities with an exponent 1 (linear scaling) and 0.6, respectively for $\tau_1$ and $\tau_2$.}
 \label{fig:fig5}
\end{figure}

To quantify the amount of spatial fluctuations at each time, we calculate the quantity  $H(t)=\int{\sigma_{OD}(x,z;t)dxdz}/\int{\overline{OD}(x,z;t)dxdz}$, where $\sigma_{OD}(x,z;t)$ is the root mean squared deviation of the ODs at time $t$ and the integration is performed over the whole image.
Figure~\ref{fig:fig5}(a) shows the amount of fluctuations $H(t)/H(0)$ as a function of time $t-t_c$ for the different low evaporation rates and highlights a clear exponential decay for all of them. 

The decay time $\tau_2$ extracted from an exponential fit is shown in \aref{fig:fig5}(b) as a function of $\tau_Q$. The linear fit to $\tau_2$ in the log-log scale corresponds to a power-law dependence as $\tau_2 \propto \tau_Q^{0.63(5)}$. This clearly differs from the linear dependence of the BEC formation time $\tau_1$ obtained in \aref{sec:Growth} [as shown in \aref{fig:fig5}(b)] and indicates that the relaxation time of the turbulence depends on $\tau_Q$ in a nontrivial way. 

\begin{figure}[t!]
 \centering
\includegraphics[width=1.0\columnwidth]{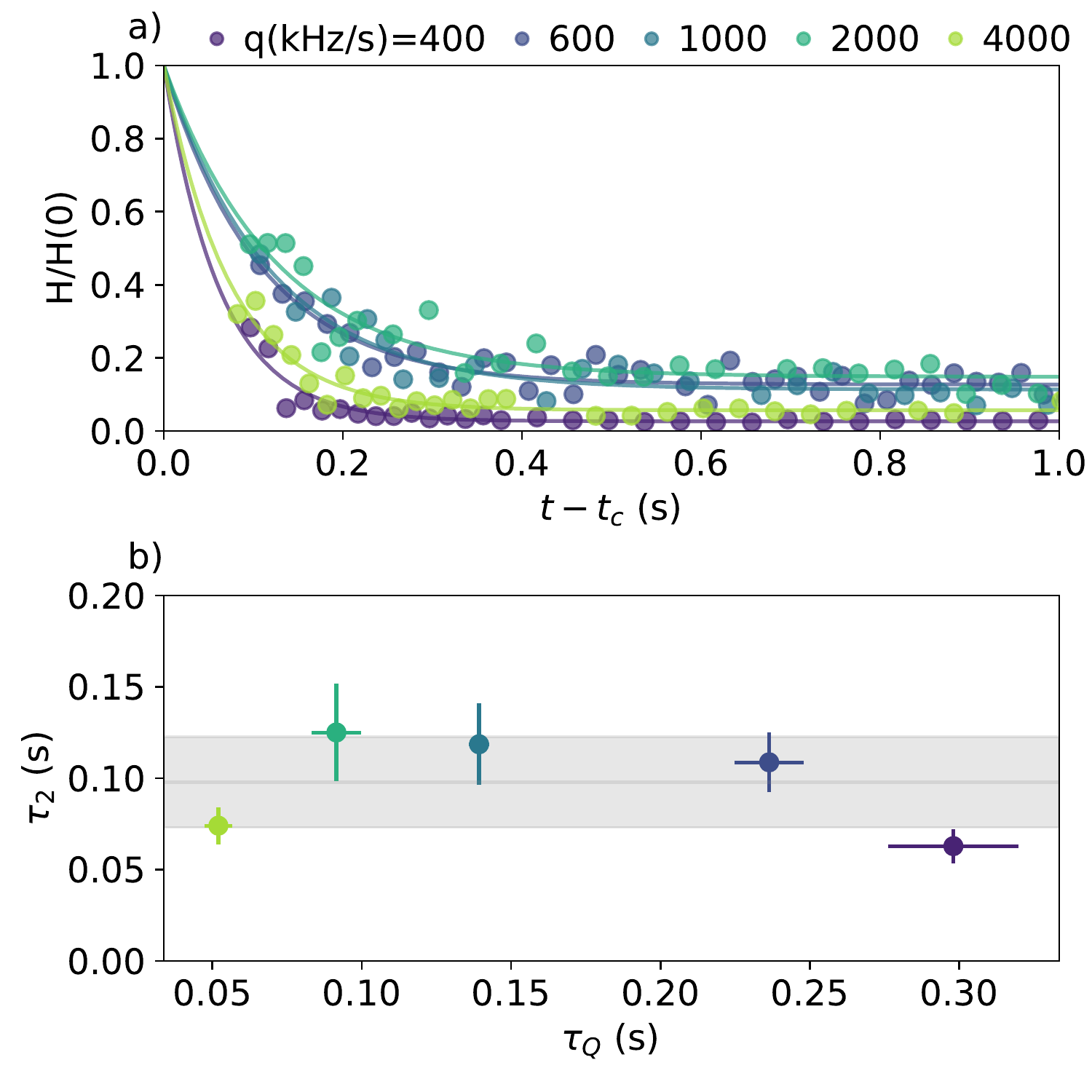}
 \caption{a) Decay of the order parameter fluctuations ${H(t-t_c)}$ in the case of high evaporation rates, normalized by the initial value $H(0)$. 
 b) Time constant $\tau_2$ extracted from the exponential fit, plotted as a function of the quench time $\tau_Q$. In contrast to the low evaporation rates, there is no clear dependence on the quench time, with $\tau_2$ resulting fairly constant at a value of \SI[separate-uncertainty=true]{98(25)}{\milli\second} (gray area).}
 \label{fig:fig6}
\end{figure}

Table \ref{table:summary} summarizes the parameters extracted from the analysis reported in this manuscript for all the combinations of atom number and evaporation rates explored. 
Note that the large atom number is particularly important for carefully studying the turbulence dynamics in the condensate. Indeed, the atom number we refer to as ``low'' is still larger than typical condensate atom numbers in the literature.

We repeated a similar analysis for the high evaporation rates ($q \geq \SI{400}{\kilo\hertz\per\second}$), where the growth of the condensate and the decay of the fluctuations occur after the end of the evaporation ramp. In this case, we  obtained that the decay time $\tau_2$ is fairly constant and on the order of \SI{100}{\milli\second}, as shown in \aref{fig:fig6}. This value is similar to the latency time, suggesting also here a connection with the collisional rate within the cloud.

\begin{table} [h!]
\resizebox{\columnwidth}{!}{%
\begin{tabular}{  c c c c c c c c}
\hline
\hline
HIGH N \\
\hline
 $q$ (kHz/s) & $t_{f}$ & $\tau_Q$  & $t_c$ & $\tau_1$  &$t_{50_\%}$ & $\tau_2$  & $\Delta t$\\ 
 \hline
20 & 11 & 18.9(1.9) & 3.19 (7)& 2.13 (5) & 4.65(4)& 0.75 (17) & -\\ 

50 & 4.4 & 7.13(18) &  1.68 (2)& 0.818 (2) & 1.877 (8) & 0.25 (2) & - \\ 

100 & 2.2 & 2.60(16) &  1.30 (1) & 0.35 (1) & 0.707 (6)& 0.14 (3) & - \\ 

200 & 1.1 & 1.23(5) &  0.70 (1) & 0.16 (1) & 0.392 (5)& - & - \\ 

300 & 0.73 & 0.59(2) &  0.531 (3)& 0.075 (2) & 0.233 (2) & 0.04 (1) & - \\ 

400 & 0.55 & 0.29(2) &  0.460 (4)& 0.030 (1) & 0.168 (2)& 0.060 (9) & 0.086 (15) \\ 

600 & 0.36 & 0.25(2) &  0.313 (3)& - & - & 0.109 (16) & 0.090 (10) \\ 

1000 & 0.22 & 0.135 (3) &  0.196 (1) & - & - & 0.14 (5) & 0.084 (10) \\ 

2000 & 0.11 &  0.104(9)  & 0.092 (9)& - & - & 0.13 (3) & 0.085 (15)\\ 

4000 & 0.055 &  0.050(4)  & 0.054 (5) & - & - & 0.075 (10) & 0.082 (17)\\ 

 \hline
 \hline
 &  &   & &  &  &  & \\ 

\end{tabular}}
\\
\resizebox{\columnwidth}{!}{%
\begin{tabular}{  c c c c c c c c}
\hline
\hline
LOW N \\ 
\hline
 $q$ (kHz/s) & $t_{f}$ & $\tau_Q$    & $t_c$  & $\tau_1$  &$t_{50_\%}$ & $\tau_2$  & $\Delta t$\\ 
 \hline
20 & 7.5 &  11.2 (6) & 3.64 (7)& 1.39 (4) & 2.95 (3)& 0.80 (11) & -\\

30 & 5 &  7.3 (5) & 2.52 (7)& 0.88 (6) & 2.04(4)& - & - \\ 

40 & 3.84 &  5.4 (4) & 2.33 (4) & 0.76 (5) & 1.63 (4)& 0.23 (4) & -\\ 

43 & 3.5 &  4.5 (3) & 2.05(4)& 0.60 (5) & 1.32 (4)& -  & -\\ 

50 & 3.0 &  4.06 (2) & 1.92 (3)& 0.52 (3) & 1.21 (3)& 0.19 (9) & -\\ 

75 & 2.0 &  2.3(3) & 1.28 (5)& 0.32 (7) & 0.87 (10)& - & -\\ 

200 & 0.96 &  0.49 (6)  & 0.661 (9)& 0.054 (3) & 0.262 (2)& - & -\\

300 & 0.50 &  0.357 (9) & 0.433 (2)& - & -& - & 0.148 (18) \\ 

750 & 0.20 &  0.17 (3) & 0.189 (4)& - & -&- & 0.142 (19)\\ 

1500 & 0.10 &  0.108 (19) & 0.102 (3)& - & -& -& 0.17 (2) \\ 

3000 & 0.05 &  0.16 (2) & 0.07 (2)&- & -& - & 0.18 (2) \\ 
 \hline
 \hline
 
\end{tabular}}

\caption{Relevant time-scales for the different ramps for $N_\text{high}$ (above) and $N_\text{low}$ (below). All time-scales are in seconds. $t_f$ is the duration of the ramp and $\Delta t$ is the latency time measured for high evaporation rates, where the condensation occurs as a post-quench process.}
\label{table:summary}

\end{table}

\section{Conclusions}
\label{sec:Conclusion}
In conclusion, by cooling a harmonically trapped bosonic gas across the critical temperature for condensation at different rates, we studied the growth of the order parameter and the temporal evolution of its spatial fluctuations. We found that the condensate starts forming after a latency time from the critical condition; this latency time inversely depends on the classical collision rate of the gas. 
If the evaporation is sufficiently fast, the condensate starts forming after the end of the evaporation ramp and the growth of the order parameter follows a universal curve with an intrinsic growth rate independent of the quench time $\tau_Q$. For slower evaporation rates, condensation occurs during the evaporation ramp and the growth curves still manifest a universal behavior, but with a characteristic timescale that linearly scales with $\tau_Q$.
We also investigated the amount of spatial fluctuations of the order parameter, formed at the early stage of condensation, and found that they decay exponentially in time and the characteristic time for relaxation obeys a power-law scaling with $\tau_Q$ with exponent $\approx 0.6$.

This work contributes to the understanding of the complex dynamics characterizing systems across phase transitions and can trigger novel theoretical investigation going beyond the current state of the art.\\

During the completion of this manuscript, we became aware of a related work \cite{Goo21b}.

\begin{acknowledgments}
We thank D. Clément and S. Giorgini for stimulating discussions.
We acknowledge funding from the European Union’s Horizon 2020 Programme through the NAQUAS project of QuantERA ERA-NET Cofund in Quantum Technologies (Grant Agreement No. 731473) and from Provincia Autonoma di Trento. 
This work was supported by Q@TN, the joint lab between University of Trento, FBK - Fondazione Bruno Kessler, INFN - National Institute for Nuclear Physics and CNR - National Research Council.

\end{acknowledgments}


%

\end{document}